\documentclass[journal=jpclcd,manuscript=article]{achemso}

\newcommand{\figwidth}{0.8 \linewidth}

\usepackage{graphicx}

\newcommand{\chtfp}{CH$_2$F$^+$}
\newcommand{\chdp}{CH$_3^+$}
\newcommand{\chdf}{CH$_3$F}
\newcommand{\tpj}{$^3P_J$}
\newcommand{\tpt}{$^3P_2$}

\newcommand{\chtfps}{CH$_2$F$^+$ }
\newcommand{\chdps}{CH$_3^+$ }
\newcommand{\chdfs}{CH$_3$F }

\newcommand{\p}{$^+$}
\newcommand{\pp}{$^+$ }
\newcommand{\fxe}{f$_{X1}$}
\newcommand{\fxt}{f$_{X2}$}
\newcommand{\fxd}{f$_{X3}$}
\newcommand{\fAe}{f$_{A1}$}
\newcommand{\fBe}{f$_{B1}$}

\title{Study of the Ne(\tpt) + \chdfs Electron Transfer Reaction below 1 Kelvin}
\author{Justin Jankunas}
\author{Benjamin Bertsche}
\author{Andreas Osterwalder}
\email{andreas.osterwalder@epfl.ch}
\phone{+41 (0)21693 79 71}
\affiliation{Institute for Chemical Sciences and Engineering, Ecole Polytechnique F\'ed\'erale de Lausanne, 1015 Lausanne, Switzerland}

\begin{document}

\begin{abstract}
Relatively little is known about the dynamics of electron transfer reactions at low collision energy. 
We present a study of Penning ionization of ground state methyl fluoride molecules by electronically excited neon atoms in the 13 $\mu$eV--4.8 meV (150 mK--56 K) collision energy range, using a neutral-neutral merged beam setup. 
Relative cross sections have been measured for three Ne($^3P_2$)+ CH$_3$F reaction channels by counting the number of CH$_3$F$^+$, CH$_2$F$^+$, and CH$_3^+$ product ions, as a function of relative velocity between the neon and methyl fluoride molecular beams. 
Experimental cross sections markedly deviate from the Langevin capture model at collision energies above 20 K. 
The branching ratios are constant. 
In other words, the chemical shape of the CH$_3$F molecule, as seen by Ne($^3P_2$) atom, appears not to change as the collision energy is varied, in contrast to related Ne($^3P_J$) + CH$_3$X (X=Cl and Br) reactions at higher collision energies.
\end{abstract}

\section{Introduction}
The experimental and theoretical study of electron transfer reactions addresses one of the most fundamental processes in chemistry and biology. 
Such reactions occupy a special place in the world of gas phase reaction dynamics because they were among the first reactions that were studied during the first phase of molecular beam based studies of reaction dynamics\cite{Levine:2009fb}. 
Penning ionization (PI), commonly described as A* + B$\rightarrow$A + B\p + e$^-$, can be viewed as an electron transfer reaction wherein the internal energy of one reactant, A*, is greater than the ionization potential of the other, B. 
The reaction proceeds by an electron transfer from target molecule B into one of the empty orbitals of A*, leading to ejection of the least tightly bound electron in species A. 
A great deal of experimental measurements and theoretical calculations have been performed on simple PI systems\cite{Siska:1993cf}, most notably atom-atom and atom-diatom reactions.
They suggest that electronic orbital overlap between the reactants determines the internal state population distribution of the B\pp product ion, as well as the Penning electron kinetic energy spectrum\cite{Niehaus:1981vb}. 
A particularly interesting case of PI is when B is a small polyatomic molecule.
Then, the presence of several (non-degenerate) molecular orbitals means that the potential energy, associated with a particular geometry of the [AB]* complex, can vary for different A--B approach directions. 
Ionization from different molecular orbitals produces different states of the molecular cation, dictated by correlation rules between the electronic states of neutral and ionized molecular species. 
Since the shape of the molecular orbitals is directly linked to the structure of the molecule, the measurement of branching ratios can serve as a stereodynamic probe of a particular PI reaction.  

Several Penning ionization studies of small polyatomic molecules by electronically excited helium and neon atoms have been performed to date. 
For example, Vecchiocattivi and co-workers examined Ne(\tpj) + H$_2$O PI  in the range of collision energies E$_\mathrm{coll}$ 40 meV--300 meV (465 K--3480 K; here, temperature is defined as T =$\mathrm{E}_\mathrm{coll}/\mathrm{k}_B$)\cite{Brunetti:2012va,Balucani:2012wv,brunetti:13}. 
The internal Ne(\tpj) energy of $\approx$16.7 eV made accessible two channels, correlating asymptotically with a molecular water cation in the ground and electronically excited states, H$_2$O\p(X $^2$B$_1$), and H$_2$O\p(A $^2$A$_1$), respectively. 
The reaction cross section was found to increase with decreasing collision energy, suggesting an attractive nature of Ne(\tpj) - H$_2$O collisions in the E$_\mathrm{coll}$ range studied. 
The relative yield of electronically excited products increased modestly with decreasing collision energy. 
Similar branching ratio behavior has been observed in the PI of N$_2$O by He* and Ne*\cite{Biondini:2005bw}.  
On the other hand, the fraction of electronically excited NH$_3$\pp product ions decreases modestly with decreasing E$_\mathrm{coll}$, as found in the study of collisions between Ne(\tpj) atoms and NH$_3$ in the range of E$_\mathrm{coll}$ 35 meV--200 meV (406 K-- 2320 K)\cite{BenArfa:1999el}.  
Penning ionization of methyl halides, CH$_3$X, is even more interesting, due to a larger number of open reactive channels. 
Photoionization studies of \chdfs have revealed five open channels at photon energies of 16.5-17 eV: production of CH$_3$F$^+$, as well as dissociative channels producing \chtfp, \chdp, CHF$^+$, and CH$_2^+$.\cite{Olney:do}
Cross sections for double dissociation have been measured to be at least 20 times smaller than the single dissociation cross sections which in turn were similar to non-dissociative photoionization.
PI studies of methyl halides have shown the dissociative channels CH$_3^+$, CH$_2$X$^+$, and the non-dissociative channel CH$_3$X$^+$.
Brunetti et al.\cite{Brunetti:1997kr} have investigated the Ne(\tpj) + CH$_3$X (X=Cl and Br) reactions and reported an increasing fraction of parent ions with decreasing collision energy, whereas the fragment yield decreased with diminishing E$_\mathrm{coll}$. 
The general assumption in the interpretation of these experiments was that the propensities to form ions in the different electronic states is related to the orientation of the molecule relative to the angle of incidence of the metastable atom.
In the case of NH$_3$+Ne($^2P_J$), for example, the ionic ground state is formed by removing an electron from the nitrogen lone pair while the first excited state is formed by removing an electron from one of the N-H bond orbitals.\cite{BenArfa:1999el}
In PI, accordingly, ground state NH$_3^+$ is formed when the Ne* approaches along the N-lone pair axis while electronically excited NH$_3^+$(A) is formed when the Ne* approaches along an N-H bond.
If the ground state cation is stable to dissociation but the excited state is not, then mass selective detection of the product ions is a direct measure of the branching ratio during the PI and thus a probe for the stereo dynamics of the reaction.
In those cases where the ionic states formed upon electron transfer can dissociate along different routes and form the same products one can make use of the fact that the internal branching ratios, i.e. those that determine the fate of ions formed in particular states, are independent of the collision energy.
Any dependence of the measured branching ratios on E$_\mathrm{coll}$ can thus be attributed to the stereo dynamics of the electron transfer process itself.
All of the PI experiments mentioned above have been carried out at collision energies close to and above $\mathrm{E}_\mathrm{coll}/\mathrm{k}_B$=300 K (25 meV). 
We report on an experimental study of Ne(\tpt) + \chdfs PI in the 13 $\mu$eV-- 4.8 meV (150 mK--56 K) collision energy range. 
The present study represents the first investigation of PI of a six-atom reaction system in this energy range.
To this date, the largest molecule studied below 10 meV was ammonia.\cite{Bertsche:2014vj,Bertsche:2014ub,jankunas:jcp}
Unlike the Penning fragment branching ratios that exhibit a noticeable dependance on E$_\mathrm{coll}$  above 300 K, we find that the relative yield of three reaction channels CH$_3$F$^+$, \chtfp , and \chdps is, within the experimental uncertainty, constant in the E$_\mathrm{coll}$ range studied.
This observation is surprising: a straightforward argument based on the well depth for different [Ne-\chdf ]* configurations, as was successfully applied in the case of Ne(\tpj) + H$_2$O reaction at higher collision energies,\cite{Brunetti:2012va} predicts a particular reaction channel to become dominant with ever decreasing collision energy. 

\section{Experimental}
The low collision energy of 13 $\mu$eV was achieved in a neutral-neutral merged beam technique.\cite{Henson:2012kr,Narevicius:2014te,Bertsche:2014vj,Bertsche:2014ub,jankunas:jcp}
The experimental apparatus is described elsewhere\cite{Bertsche:2014ub,jankunas:jcp}. 
Briefly, a supersonic beam of neon atoms in the excited Ne(2p$^5$ 3s$^1$, \tpt) state is bent by means of a magnetic hexapole guide, and merged with a molecular beam of electrostatically guided \chdfs molecules in the electronic ground state.\cite{Rangwala:2003cy,Sommer:2009cc,Bertsche:2011jr}
Only the $J=2$ spin-orbit component of the 2p$^5$ 3s$^1$ $^3P_J$ state of neon is present in the reaction since the $J=1$ component is short-lived and the $J=0$ component is not paramagnetic.
The rotational state distribution of CH$_3$F molecules is broader and harder to quantify than the single-state Ne beam. 
Based on resonance enhanced multiphoton ionization (REMPI) spectra of ND$_3$ molecules that are co-expanded and guided together with the desired CH$_3$F molecules we estimate the rotational temperature to be $\approx$ 20 K.
Because of the rotational state dependent guiding dynamics the rotational degrees of freedom are not in thermal equilibrium.\cite{Bertsche:2011jr}
Calculating the CH$_3$F (J, K) state populations present in the interaction region is challenging. 
Based on our earlier finding wherein ND$_3$ molecules before entering and after exiting the molecular guide exhibited different rotational temperatures,\cite{Bertsche:2011jr} we must assume that rotational distributions of CH$_3$F molecules before and after entering the electric field region are also different.
Assuming T$_{rot}$ = 20 K for CH$_3$F molecules immediately following the supersonic expansion, then over 95$\%$ of molecules reside in J $\leq$ 5 levels. 
More importantly, over 60$\%$ of CH$_3$F molecules occupy the K = 0 state, whilst the remaining $\approx$ 40$\%$ fraction of molecules are in 3 $\leq$ K $\leq$ 1 states. 
Methyl fluoride molecules in J $\geq$ 1, K = 0 rotational levels exhibit only a quadratic Stark effect, and are therefore expected to be guided less efficiently than the K $\geq$ 1 states whose energy is linearly proportional to the electric field strength inside the hexapole guide. 

The crossing angle between the two merged molecular beams is zero, and the average relative velocity between reactants v$_\mathrm{rel}$ is zero when the laboratory speeds of Ne(\tpt) and \chdfs beams are the same.  
As explained previously\cite{Bertsche:2014vj,jankunas:jcp,Shagam:2013ev}, the resolution in E$_\mathrm{coll}$ is dictated mainly by a finite duration of the molecular beam pulse.
In the present case the velocity spread at each relative velocity is $\approx$ 28 m/s.
The lowest attainable collision energy is half of that because with both beams centred around the same velocity the average relative velocity is $<$v$_{rel}>$=0$\pm$28 m/s, and $<|$v$_{rel}|>$=14$\pm$14 m/s, assuming a square distribution of the velocities.
The spread of collision energies scales quadratically with the average velocity, and the minimum central velocity we can reach corresponds to an average collision energy E$_\mathrm{coll, min}/\mathrm{k}_B$ = 120 mK. 
Transverse velocities in the present experiment contribute to the spread in collision energy to a lesser amount and are neglected.\cite{jankunas:jcp}
Fragment ions are collected in a time-of-flight mass spectrometer and counted mass-selectively.
The number of collected ions is proportional to the absolute reaction cross section and the beam--densities and --overlap; due to the difficulties to quantify the latter two, no absolute cross sections can be reported.    

\section{Results and Discussion}
\begin{figure}
 \includegraphics[width=\figwidth]{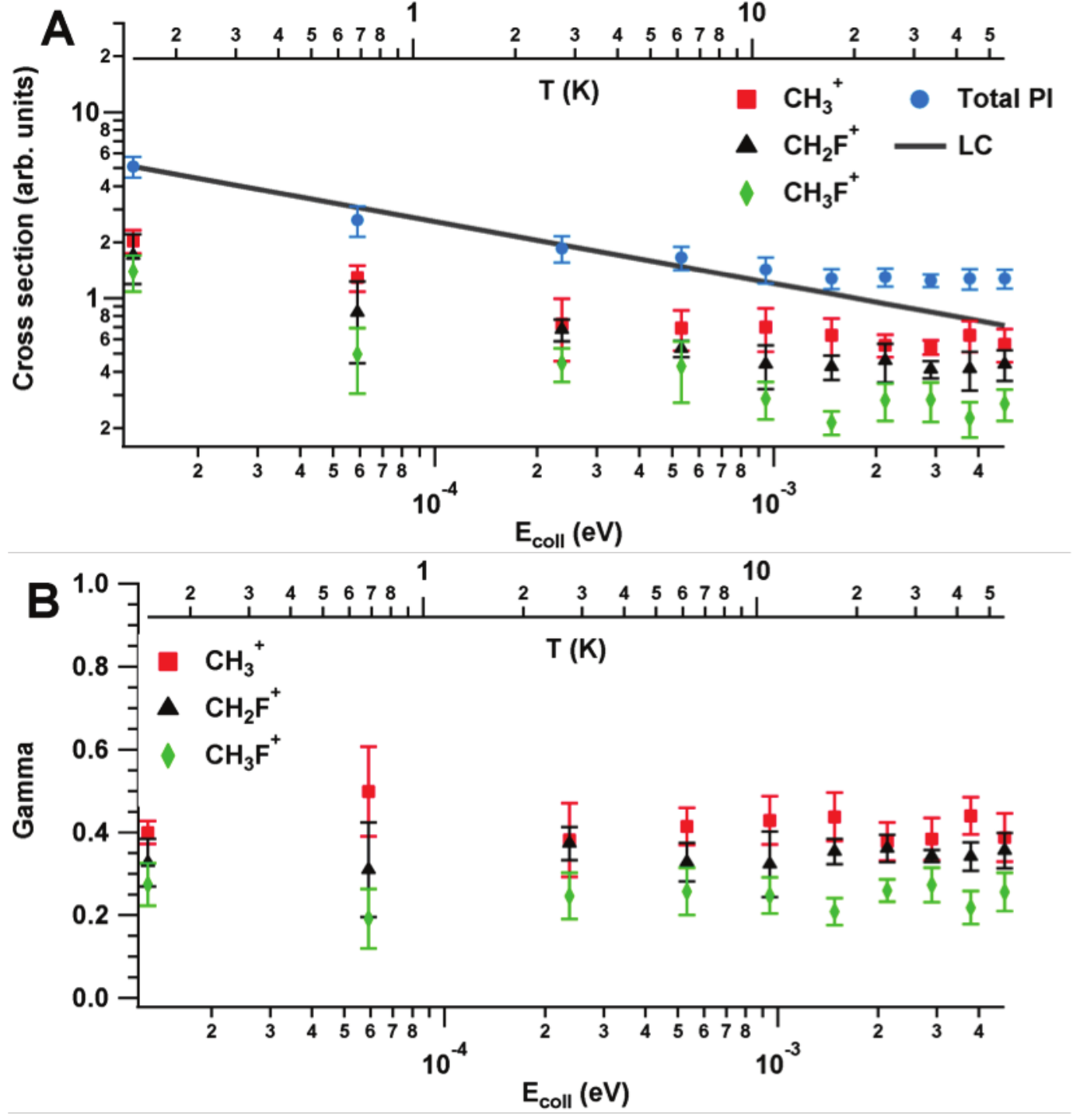}%
\caption{\label{fig:data}(A) Partial and total Ne(\tpt) + \chdfs Penning ionization cross sections. Red squares: \chdps products; black triangles: \chtfps products; green rhombs: CH$_3$F$^+$ products; blue circles: total PI cross section (\chdp + \chtfp + CH$_3$F$^+$). The solid grey curve shows the prediction based on the Langevin capture model. (B) Branching ratios, $\Gamma$(i), for the three reactive channels as a function of collision energy. Color coding is the same as in panel (A).}
 \end{figure}

Measured reaction cross sections are shown in \ref{fig:data}A. 
Cross sections for \chdps (red squares), \chtfps (black triangles), and CH$_3$F$^+$ (green rhombs) product formation are shown individually, and the sum is plotted as blue circles, showing the total Ne(\tpt) + \chdfs PI probability. 
The grey curve is the Langevin capture (LC) model prediction for an intermolecular potential V(R) $\propto$ R$^{- 6}$, where the cross section $\sigma\propto\mathrm{E}_\mathrm{coll}^{-1/3}$. 
The LC model predicts $\sigma$ to decrease with increasing collision energy.
A log-log plot of $\sigma$ vs. E$_\mathrm{coll}$ yields a straight line with a slope of -1/3\cite{Niehaus:1981vb}. 
The Ne(\tpt) + \chdfs molecular collisions are dominated by long range interactions proportional to V(R) $\propto$ R$^{- 6}$ up to about 20 K. 
At E$_\mathrm{coll}>$ 20 K the data start do deviate from the LC model, suggesting that other potential energy terms begin to contribute.      

\begin{figure}
 \includegraphics[width=\figwidth]{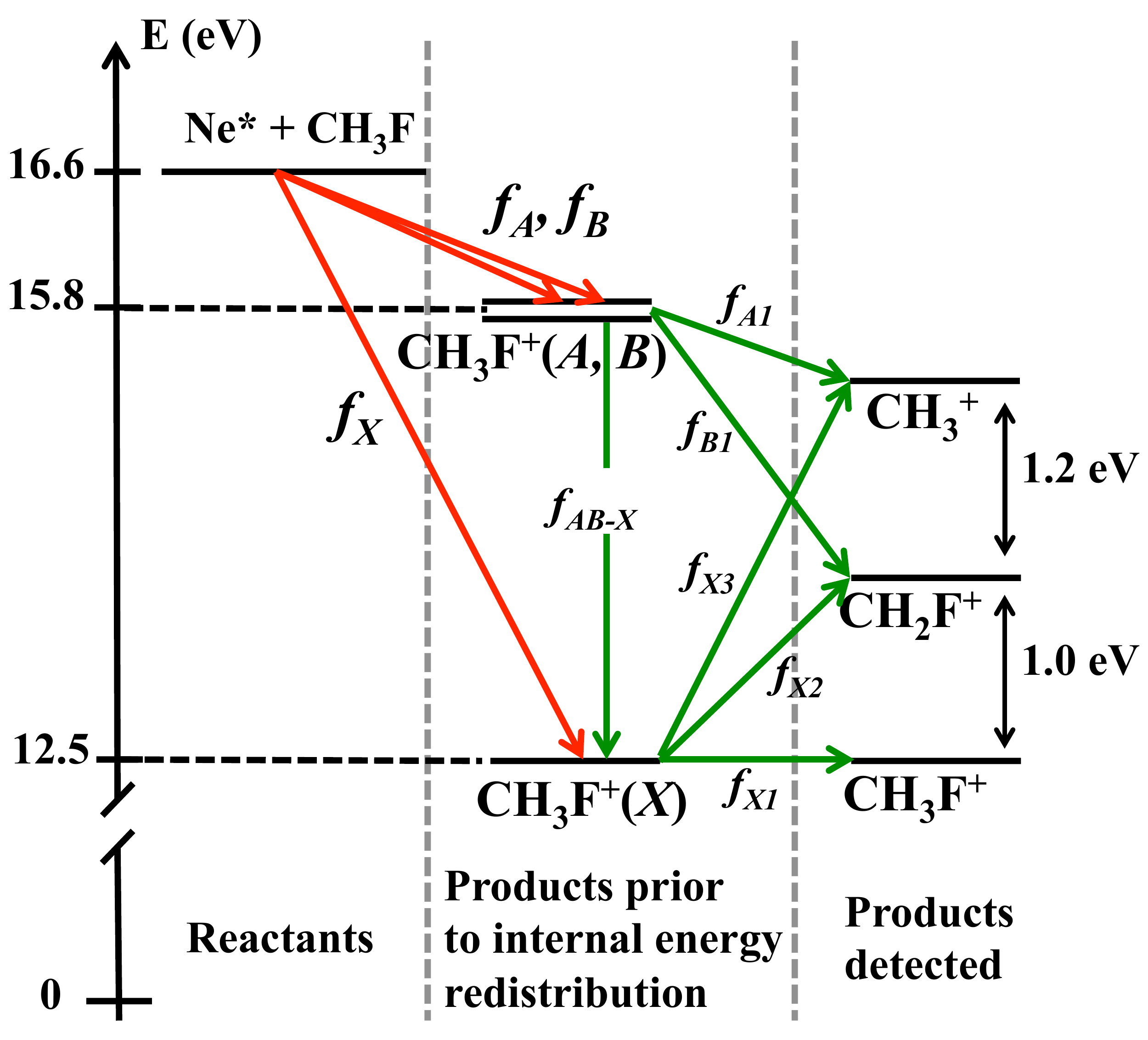}%
\caption{\label{fig:scheme}Energy diagram for the Ne(\tpt) + \chdfs reaction, illustrating the three accessible electronic states of CH$_3$F$^+$ product cation, with the associated collision induced and internal conversion/dissociation branching ratios indicated by red and green arrows, respectively.}
 \end{figure}

We now turn to the measured branching ratios.
Experimental results, defined as 
\begin{equation}
\Gamma(i) = \frac{[\mathrm{i}]}{\mathrm{[CH_3^+ ] + [CH_2F^+] + [CH_3F^+]}},\label{eq:ratio}
\end{equation}
for the three reaction channels are shown in \ref{fig:data}B.

In order to disentangle the different reaction channels that can lead to the observed branching ratios we refer to the diagram in \ref{fig:scheme}.
The left side shows the initial level of the combined Ne(\tpt)+\chdfs system at the internal energy of Ne(\tpt).
PI can lead to the formation of CH$_3$F$^+$ ions in the X, A, or B state, as shown in the central part.
These states can decay into the observed fragments on the right side of \ref{fig:scheme}.
The experimentally observed branching ratios are a convolution of the desired branching ratios f$_X$, f$_A$, and f$_B$ and the  internal branching ratios, labeled as  \fxe, \fxt, \fxd, f$_{AB-X}$ \fAe, and \fBe, which are not known.
The excited A and B states have short life times such that it can be assumed that all excited ions will decay by one of the pathways indicated in the figure.
The branching ratios are related by an underdetermined set of equations.
However, limits can be set on f$_X$ that allow to shed some light on the observed dynamics.
It is known that CH$_3$X\pp cations (X = F, Cl, Br, I) in several electronically excited states are subject to various internal conversion, conical intersection, and dissociative processes.\cite{lane:93}  
However, it has been observed that in CH$_3$F\pp the only decay channel for the A and B states is the formation of \chdp.\cite{eland:76}
This implicitly means that f$_{AB-X}$ and f$_{B1}$ are all zero.
The latter can be rationalised by the coupling of the A-- and B--states.
Unlike in CH$_3$Cl\p, CH$_3$Br\p, and CH$_3$I\pp molecules, the excited A-- and B--states of the CH$_3$F$^+$ cation are degenerate. 
This makes the following analysis easier, because the internal conversion (IC) from B to A is often said to proceed with a 100\% efficiency.\cite{eland:76}
Assuming IC from B to A to be much faster than the molecular dissociation CH$_3$F$^+$(B) $\rightarrow$ \chtfps + H  (see \ref{fig:scheme}), the resulting equations can be solved to put limits on the f$_X$, f$_A$, and f$_B$ values. 
f$_{AB-X}\neq0$ would potentially open a path to CH$_3$F$^+$ formation via PI to one of the excited states.
Branching ratios for the formation of [\chdp], [\chtfp], and [CH$_3$F$^+$] are defined by equation \ref{eq:ratio}.
The detailed balance dictates the following relations between these and the individual branching ratios f$_X$, f$_A$, f$_B$, and f$_{X1}$, f$_{X2}$, f$_{X3}$, f$_{A1}$, f$_{B1}$ (see \ref{fig:scheme}):
\begin{eqnarray*}
\Gamma(\mathrm{CH_3^+}) &=& f_Xf_{X3} + f_Af_{A1}\\
\Gamma(\mathrm{CH_2F^+}) &=& f_Xf_{X2} + f_Bf_{B1}\\
\Gamma(\mathrm{CH_3F^+}) &=& f_Xf_{X1}\\
f_X + f_A + f_B &=& 1\\
f_{X1} + f_{X2} + f_{X3} &=&1\\
f_{A1} + f_{B1} &=& 1.
\end{eqnarray*}
The assumption that the IC is much faster than the corresponding CH$_3$F$^+$(B) $\rightarrow $\chtfp + H dissociation, allows the above equations to be simplified because f$_{B1} $= 0. 
Furthermore, f$_{AB-X}$ has been observed to be zero\cite{eland:76} by the lack of CH$_3$F$^+$ formation via the excited A- or B-states.
Making use of the fact that all $\Gamma$[i] are, within the experimental uncertainty, insensitive to the collision energy we can average the branching ratios over the entire collision energy range studied to obtain $\Gamma$(\chdp ) = 0.42 $\pm$ 0.04, $\Gamma$(\chtfp ) = 0.34 $\pm$ 0.02, and $\Gamma$(CH$_3$F$^+$) = 0.24 $\pm$ 0.03.
Limits on f$_X$ are then given as 0.58 $<$ f$_X$ $<$ 1.00 by assuming either f$_A*$\fAe=0, in which case f$_X$=1, or by assuming \fxd=0 in which case f$_X$=$\Gamma$(CH$_3$F$^+$)+$\Gamma$(\chtfp ) = 0.58 and f$_A$+f$_B$=0.42.
In the limit f$_X$ = 1 we have, by definition, f$_B$ =f$_A$ = 0.
The resulting CH$_3$F$^+$(X) cation dissociates into experimentally detected CH$_3$F$^+$, \chtfp , and \chdps products with branching ratios of \fxe = 0.24, \fxt = 0.34, and \fxd = 0.42, respectively. 
Conversely, if f$_X$ = 0.58, then 42\% of all Ne(\tpt) + \chdfs collisions populate the degenerate A and B electronic states which, assuming fast IC (\emph{vide supra}), dissociate completely into \chdps products. 
In this scenario the CH$_3$F$^+$(X) products branch out into the detected CH$_3$F$^+$ and \chtfps ions with f$_{X1}$ = 0.41, and \fxt= 0.59, respectively, whilst \fxd = 0.     

Given the energetics of the reaction it is safe to assume the \fxe, \fxt, \fxd, \fAe, and f$_{B1}$ to be independent of collision energy.
Consequently, also the f$_X$, f$_A$, and f$_B$ remain constant as the collision energy increases from 150 mK up to 56 K which in view of previous observations at higher E$_\mathrm{coll}$ is surprising.
It is tempting to attribute this finding solely to the fact that the current experiment was performed at low collision energies, whereas previous branching ratios that did depend on E$_\mathrm{coll}$, were measured at E$_\mathrm{coll}$ $>$ 25 meV (300 K). 
The increasing fraction of CH$_2$X\pp (X = Cl, Br) and \chdps Penning fragments with growing collision energy has been explained by drawing attention to different potential energy slopes for Ne(\tpj) approach toward the -X and -CH$_3$ end of the CH$_3$X molecule.\cite{Brunetti:1997kr}
They suggest that the softer repulsive potential around the methyl group results in a greater fragmentation at higher E$_\mathrm{coll}$. 
While it is hard to rule out the above argument based on the current findings at E$_\mathrm{coll}$ $<<$ 300 K, one would expect a particular reaction channel, in this case Ne(\tpj) approach from either side along the C-F axis of \chdf, to be ever more favored as E$_\mathrm{coll}$ $\rightarrow$ 0 eV. 
This is not what is observed experimentally. 
One way to understand the constant yield of the three reaction products as a function of collision energy is to consider the f$_X$ = 1 scenario:
all reactive Ne(\tpt) + \chdfs collisions populate only the ground electronic state of CH$_3$F$^+$(X) ion which then dissociates into the detected products with the branching ratios \fxe, \fxt, and \fxd.
This hypothesis is disfavored by the experiments wherein Penning electrons, as opposed to fragment ions, have been detected. 
Methyl fluoride molecules, ionized with He I and Ne I radiation, yielded electrons with a bimodal kinetic energy distribution, corresponding to the production of CH$_3$F$^+$(X) and CH$_3$F$^+$(A, B) ions.\cite{eland:76} 
A related Ne(\tpj) + NH$_3$ system correlating with NH$_3$\pp and NH$_2$\pp product ions has been shown to proceed via NH$_3$\p(X) and NH$_3$\p(A) electronic states, respectively, as deduced from electron kinetic energy analysis.\cite{BenArfa:1999el}
The exclusive formation of ground state CH$_3$F$^+$(X) product ions is therefore unlikely, although not impossible.

\section{Conclusions}
Study of electron transfer reactions at collision energies below 1 Kelvin (85 $\mu$eV) has only begun to unravel our limited understanding of low-energy molecular collisions. 
In addition to the observation of elusive shape resonances in the He* + H$_2$ PI,\cite{Henson:2012kr,Narevicius:2014te} chemical reactions at low temperatures exhibit rather unusual stereodynamics.\cite{deMiranda:2011gd}
The study of the Ne(\tpt) + \chdfs reaction in the low temperature regime has shown yet again energy independent branching ratios, similar to what has been observed in Ne(\tpt)+ammonia.\cite{Bertsche:2014vj,Bertsche:2014ub,jankunas:jcp}
Similar Penning ionization studies carried out at collision energies close to and above room temperature have shown that branching ratios do depend on E$_\mathrm{coll}$.
In certain cases the ratio of dissociative PI increases with a growing collision energy\cite{BenArfa:1999el,Brunetti:1997kr}, and at other times it does the opposite.\cite{Brunetti:2012va,Balucani:2012wv,brunetti:13,Biondini:2005bw}
We have not been able to conclusively establish the origin of the constant $\Gamma$(i) values, and we are left to speculate:
(a) The dynamics of Ne(\tpt) + \chdfs reaction at low temperatures changes drastically, and only the CH$_3$F$^+$(X) product ion in its ground electronic state is produced. 
This would be in sharp contrast to PI experiments at higher energies, where several (energetically accessible) electronic states of the polyatomic cation are populated. 
(b) A more probable and, unfortunately, currently less insightful, scenario is the combination of the increased de Broglie wavelength, with the ratio of collision time to rotational period of \chdf, which also increases with decreasing E$_\mathrm{coll}$.
Although a quantitative assessment of this effect at this point is not possible it is easily imagined that this also results in constant branching ratios. 
Fully dimensional Ne(\tpt) + \chdfs quantum calculations at the moment appear to be unaffordable, however classical or quasiclassical calculations could potentially shed some light on the question of constant branching ratios in Ne(\tpt) + \chdfs reaction. 
(c) Finally it must also be pointed out that our energy-dependent branching ratio expectations are based on high temperature reaction dynamics between reagents in the ground electronic states. 
Often such reactions have significant (chemical) reaction barriers. 
The height of the barrier depends on the direction of approach of the reactants - a realm of  stereodynamics, which taught us the idea of a minimum energy path. 
Barrierless reactions, like Ne(\tpj) + \chdf, that occur at low collision energies have a unit opacity function at large internuclear separations.\cite{Niehaus:1981vb} 
Theoretical calculations are once again needed to provide a full understanding by calculating, for example, the energy difference between several [Ne-\chdf ]* conformers at large internuclear separations. 

\begin{acknowledgement}
We thank Rainer Beck and Marcel Drabbels (both EPFL) for lending us scientific equipment and the Swiss Science foundation (grant number P00P2-144924) and the EPFL for funding. 
\end{acknowledgement}

\providecommand{\latin}[1]{#1}
\providecommand*\mcitethebibliography{\thebibliography}
\csname @ifundefined\endcsname{endmcitethebibliography}
  {\let\endmcitethebibliography\endthebibliography}{}

\end{document}